# MAP PROJECTIONS MINIMIZING DISTANCE ERRORS

J. Richard Gott, III[1], Charles Mugnolo[1], and Wesley N. Colley[2]
[1]Princeton University Department of Astrophysical Sciences, Peyton Hall, Ivy Lane, Princeton, NJ 08544 (jrg@astro.princeton.edu)
[2]Center for Modeling, Simulation and Analysis, University of Alabama in Huntsville, D-15 Von Braun Research Hall, Huntsville, AL 35899 (colleyw@uah.edu)

**ABSTRACT**
It is useful to have mathematical criteria for evaluating errors in map projections. The Chebyshev criterion for minimizing rms (root mean square) local scale factor errors for conformal maps has been useful in developing conformal map projections of continents. However, any local error criterion will be minimized ultimately by map projections with multiple interruptions. A disadvantage of interrupted projections is that some pairs of points that are close on the globe are far apart on the map. Since it is as bad to have two points on the map at two times their proper separation as to have them at half their proper separation, it is the rms logarithmic distance between random points in the mapped region that we will minimize. The best previously known projection of the entire sphere for distances is the Lambert equal-area azimuthal with an rms logarithmic distance error of $\sigma = 0.343$. For comparison, the Mercator has $\sigma = 0.444$, and the Mollweide has $\sigma = 0.390$. We present new projections: the "Gott equal-area elliptical" with perfect shapes on the central meridian, the "Gott-Mugnolo equal-area elliptical" and the "Gott-Mugnolo azimuthal" with rms logarithmic distance errors of $\sigma = 0.365$, $\sigma = 0.348$, and $\sigma = 0.341$ respectively, which improve on previous projections of their type. The "Gott-Mugnolo azimuthal" has the lowest distance errors of any map and is produced by a new technique using "forces" between pairs of points on a map which make them move so as to minimize $\sigma$. The "Gott equal-area elliptical" projection produces a particularly attractive map of Mars, and the "Gott-Mugnolo azimuthal" projection produces an interesting map of the moon, both of which we show.

## 1 Distance Errors

Various authors have developed mathematical criteria for judging errors in map projections. Airy (1861) produced an azimuthal projection which minimized the sum of the squares of the errors in local scale perpendicular and parallel to the radii. The Chebyschev (1856) criterion for minimizing the rms (root-mean-square) scale factor errors in a conformal map of a finite region of the sphere showed that the best such map is one where the scale factor along the boundary is a constant. This criterion has been applied by Mitchell Feigenbaum in the Hammond Atlas (1992) to produce improved conformal maps of individual continents, where the shape of the boundary follows the boundary of the continent (c.f. Mundell 1993 for discussion). For a spherical cap the best conformal map by this criterion is the stereographic projection. The Eisenlor projection with a boundary cut along the 180º meridian of longitude fulfills this criterion, with a constant scale factor on the boundary cut which is a factor of $3+\sqrt{2}$ larger than at the center of the map (0º longitude and latitude). However, it is not clear that this is the best conformal map of the globe. For example, if one extended the boundary cut to also include the prime meridian, the map would be separated into two hemispheres sitting side by side and these would be shown with the stereographic projection with two both hemispheres so that the scale at the boundary is only a factor of 2 larger than at the center of each hemisphere. This would have smaller rms local scale errors than the Eisenlohr projection. Any local criterion involving scale factor or isotropy (Tissot ellipses) will obviously be improved by just adding more boundary cuts. Indeed Goode (1925) has advocated this, making interrupted equal-area projections with boundary cuts in the oceans to produce smaller errors in local isotropy. Buckminster Fuller (1943) produced a map based on an icosahedron where twenty triangular faces were unfolded to make a flat map—which he claimed was the most accurate map of the earth so far. This map of course had numerous boundary cuts.

But there are problems with boundary cuts. Boundary cuts interrupt the geodesics connecting two random points on the globe interfering with the continuity of the topology of features on the map. But also, importantly, boundary cuts produce global distance errors, taking two points that are close on the globe (just on opposite sides of the boundary cut) and putting them far apart on the map. In evaluating distance errors, it is just as bad if two points are portrayed at twice their proper separation on the map as if they are portrayed at half their proper separation. Therefore it is the rms logarithmic distance error between random pairs of points on the sphere that we will be interested in minimizing:

$$D = \left\langle \left[\ln(d_{map}/d_{globe})\right]^2 \right\rangle^{1/2}, \tag{1}$$

where the scale of the map projection (i.e. the size of the globe) is adjusted to minimize $D$. This is easy to evaluate in practice. Consider a large number of random pairs of points on the globe (typically we will be using 30,000,000 random pairs). Imagine a globe with the scale in miles per inch that is close to that desired for the map. For each pair we compute $x_i = \ln(d_{map,i}/d_{globe,i})$. Then we compute the mean and standard deviation of the $x_i$'s: $\mu = \langle x \rangle, \sigma = \langle (x_i - \mu)^2 \rangle^{1/2}$. Then we just multiply the size of the globe by a factor $f$ where $\ln f = \langle x \rangle$ so that with the new globe $\mu = 0$. This sets the scale of the map—which is then equal to the scale of that globe relative to the earth itself, and does not affect $\sigma$. After the scale is set we could draw a scale bar at the bottom of the map to use in measuring the distances between pairs of points on the map. It will have the property that it minimizes the rms logarithmic distance errors for random points on the map. After the scale is set, by adjusting the size of the globe, $\langle x \rangle = 0$ so $\sigma = D$. $D$ is a global quantity which is affected by boundary cuts and is never zero. It weights each pair of points equally regardless of their separation.

This we would claim improves on previous distance error measures. Gilbert (1974) proposed minimizing $D'^2 = \left\langle (d_{map} - d_{globe})^2 \right\rangle / \left[\langle d_{map}^2 \rangle \langle d_{globe}^2 \rangle\right]^{1/2}$. This measure (again using random pairs of points) has the difficulty that it weights more heavily pairs of points that are far apart on the map and on the globe and does not treat scale factor errors of constant factor (either $2x$ or $x/2$) equally. Also, when Gilbert's measure is minimized for an azimuthal projection for the globe it produces a double valued map which our techniques do not produce. Boundary cuts cause our rms logarithmic distance errors to increase but not blow up, and we can compare all pairs of points with no distance limits. (Laskokski [1997a,b] has used Gilbert's distance error measure together with errors in the local Tissot ellipse $E'^2 = \langle [(a/b)-1]^2 \rangle$, and $A'^2 = \langle [ab-1]^2 \rangle$ to produce a tri-optimal projection that minimized the sum of these three values relative to their values for the Equirectangular projection assuming a boundary cut only along the 180° meridian of longitude.)

## 2  Distance Errors in Various Map Projections

Two of us (JRG and CM) have measured the distance errors in different map projections by placing 30,000,000 random pairs of points on the globe and measuring the rms logarithmic distance error ($\sigma$) between these pairs of points on the map versus the distance between these pairs of points on the globe. (In each case, the overall size of the map relative to the map scale is adjusted to minimize the rms scale error.) See Table 1. This table includes numerous previously known projections plus some new ones we will be discussing shortly. They are in rank order by their distance errors. Projections that are equal-area are designated (EA) while conformal projections are designated (C). For compactness, in what appears below we will use the phrase "distance errors" to mean rms logarithmic distance errors.

As the table indicates, the Mercator projection has $\sigma = 0.444$. Not surprisingly, it is not very good at distances. By comparison, the Mollweide projection is better, with distance errors of $\sigma = 0.390$. It avoids the overly large polar areas of the Mercator and plots the north and south poles as points. The Winkel-Tripel projection currently used by the National Geographic Society for its world maps has distance errors of $\sigma = 0.412$. The Hammer equal-area projection is even better than the Mollweide with distance errors of $\sigma = 0.388$. Can we do better?

The Hammer equal-area projection is produced by collapsing longitudes by a factor of two so that the whole globe is mapped onto a hemisphere of the globe. This hemisphere is then mapped onto a circular disk with the Lambert equal-area azimuthal projection. Then this disk is stretched by a factor of two in the horizontal direction to produce an ellipse with an axis ratio of 2:1. Since all these transformations preserve relative areas, the final map is an equal-area map. The north pole is at the top, the south pole is at the bottom, and the equator is a straight line across the middle.

### 2.1  *Gott Equal-Area Elliptical Projection*

The new Gott equal-area elliptical projection, designed by JRG to lessen distance errors, is produced in the following way: Collapse the longitudes by a factor of two toward the central meridian so the whole globe is mapped onto one hemisphere of the globe. Now establish east and west "poles" 180° apart on the equator, at the eastern and western edges of this hemisphere, and define "new longitude and latitude" relative to these two poles. Then collapse the "new longitudes" by a factor of two (toward the equator) so that the whole globe is mapped onto a quadrant of the sphere. The north pole is now plotted at longitude 0° and latitude +45°, while the south pole is plotted at longitude 0° and latitude –45°. Both the first compression in longitude and the second compression in "new longitude" preserve relative areas, so the combination does as well. Then map this quadrant of the sphere onto a plane with a transverse equal-area Bromley-Mollweide projection. (The Bromley-Mollweide projection is like the Mollweide projection—elliptical, with elliptical longitude lines and straight latitude lines, but stretched to produce an ellipse with an axis ratio of $\pi^2/4:1$ so that the equator becomes a standard parallel where shapes are preserved locally). Since the quadrant being mapped is bounded by two lines of "new longitude" and such lines are plotted as ellipses by the transverse Bromley-Mollweide projection, the Gott projection will map the earth onto an ellipse. Since the longitude and "new longitude" compressions are by a factor of two each in the horizontal and vertical directions along the central meridian in the map and the transverse Bromley-Mollweide projection preserves shapes along this line, the Gott projection will preserve shapes locally along the central meridian. The map is an attractive ellipse with an axis ratio of $16/\pi^2:1$ or 1.62211..:1, close to the golden mean, which is $(1+\sqrt{5})/2:1$ or about 1.618:1. See Figure 1. (Earth maps are made using a C++-code developed by WNC; the source image is from Stöckli, R. [2006].) The scale is uniform and shapes are preserved locally along the central meridian. This makes a rounder map which has better distances, than the Hammer or Mollweide, but still has good shapes in the center.

The formulas for the Gott equal-area elliptical projection are as follows. Cartesian map coordinates $(x,y)$ may be calculated from the latitude and longitude $(\phi,\lambda)$ (in radians) of a point on the globe by first defining the "new latitude" and "new longitude":

$$\phi' = \sin^{-1}[\cos\phi \sin(\lambda/2)]$$
$$\lambda' = \tfrac{1}{2}\sin^{-1}[\sin\phi/\cos\phi'] \quad (2)$$

Then,

$$x = \sqrt{2}\sin\theta$$
$$y = \frac{\pi}{2\sqrt{2}}\lambda'\cos\theta \quad (3)$$
$$2\theta + \sin(2\theta) = \pi\sin\phi'$$

The Gott equal-area elliptical projection has distance errors of $\sigma = 0.365$ as compared with the Mollweide equal-area projection which has distance errors of $\sigma = 0.390$. The Mollweide projection has perfect shapes locally at only two points on the central meridian of the map, while the Gott elliptical projection has perfect shapes locally along the entire central meridian. For comparison, the Bromley-Mollweide projection which has perfect shapes locally along the equator has distance errors of ($\sigma = 0.420$)—(it is even more elongated than the Mollweide), the Hammer equal-area projection (with distance errors of $\sigma = 0.388$) has perfect shapes locally only at one point in the center of the map. The Breisemeister with $\sigma = 0.372$ has perfect shapes at only two points on the central meridian. The Eckert VI equal-area projection has distance errors of $\sigma = 0.385$, and the sinusoidal equal-area projection has distance errors of $\sigma = 0.407$. Thus, the new Gott equal-area elliptical has smaller distance errors than these other standard projections and has some nice properties in addition. It has particularly good shapes for Europe, Africa, and Antarctica which occupy the visual center of the map. Antarctica is usually portrayed rather poorly on such maps—often stretched out unrecognizably. This makes the Gott equal-area elliptical a particularly good map projection for Mars, where the polar caps are of particular interest and they as well as other features can be portrayed well in a clear and pleasant shape. See Figure 2 (source imagery JPL [2006]).

The distance errors for single hemisphere maps are given in Table 2. The distance errors here are much smaller than for the whole sphere, except for the gnomonic projection. The best projection for distances for a single hemisphere is the Lambert azimuthal equal-area which has $\sigma = 0.0829$. Snyder (1993, p. 173) remarks that it is "generally the basis for the few remaining atlas maps of the eastern and western hemispheres." The Lambert equal-area azimuthal also gives the best distances for the whole sphere displayed as two hemispheres side by side ($\sigma = 0.425$).

## 3  Improving Distance Errors

The distance errors in the Gott projection can be further minimized by compressing it horizontally until its axis ratio is 1.18:1. Compression horizontally helps with distances by bringing points on opposite sides of the boundary cut closer together. Also it makes the longitude lines from pole to pole more nearly equal in length. Gott and Mugnolo found this to be the best compression ratio (to three significant figures) to minimize distance errors. This produces the Gott-Mugnolo equal-area elliptical projection where the distance errors are only $\sigma = 0.348$. (The corresponding earth map with this projection is shown in Figure 3.) This projection has the smallest distance errors of any map projection we have found so far where the north pole is at the top and the south pole is at the bottom. It pulls points like Tokyo and Honolulu closer together improving their distance error, while increasing slightly the distance errors within places like Africa where shape compression makes east-west distances too small and north-south distances too great. Still, the Tissot ellipse at the center of the map ($0^o$ latitude, $0^o$ longitude) has $a/b = 1.37$ which is not as bad as the $a/b = 2$ found at the same location in the Gall-Peters Projection, but slightly worse than the Mollweide projection which has $a/b = 1.23$ there.

The Briesemeister (axis ratio 1.75:1) projection is better at distances than the Hammer (axis ratio 2:1). Optimally compressed versions of the Hammer (1.22:1, $\sigma = 0.352$), Mollweide (1.27:1, $\sigma = 0.359$), Aitoff (1.23:1, $\sigma = 0.354$), Eckert VI (0.67:1, $\sigma = 0.361$), Eckert IV (1.30:1, $\sigma = 0.365$) are not as good as the Gott-Mugnolo equal-area Elliptical. Furthermore, the optimally compressed Hammer and Mollweide projections have $a/b$ values at ($0^o,0^o$) of 1.64, and 1.94 respectively which give worse isotropy at the center than the Gott-Mugnolo Elliptical.

The best conformal map projection we have found so far in terms of distances is the Lagrange projection with $n = 0.47$ (accurate to two significant figures), which has distance errors of $\sigma = 0.432$. This is constructed by making a Mercator map of the globe, multiplying its scale by a factor of $n = 0.47$ and plotting it back on the globe with the inverse of the original Mercator projection. Thus, the entire globe is now conformally plotted between longitude $0°$ and $169.2° = (0.47 \times 360°)$. One then makes a conformal stereographic projection of this centered on the equator at longitude $84.6°$. Meridians of longitude and circles of latitude are arcs of circles. The angle at the north pole is $169.2°$. But the curve of distance errors versus n is very shallow and to three significant figures (near our measuring accuracy) the distance errors on this best conformal map projection are not appreciably less than those for the simple Lagrange projection (actually first invented by Lambert) with $n = 0.5$ (which has $\sigma = 0.432$). This map is produced by making a Mercator map of the globe, multiplying by a factor of 0.5 plotting it with an inverse Mercator projection back on a hemisphere of the globe and plotting this with a stereographic projection to make a circular map with the north pole at the top, the south pole at the bottom and the equator as a horizontal diameter of the circle. (The orthographic Lagrange [similar to a projection developed by Snyder] in Table 1 is produced by making an orthographic projection of this hemisphere instead.) The Lagrange map is conformal, has a nice circular shape, and for all practical purposes should be the conformal map of choice if one wants the best distances. (We were not able to check the Eisenlohr projection since it's formulas were not readily available but it seems clear that since the optimal value of $n$ was not greater than 0.5 that it would not beat the simple Lagrange with $n = 0.5$ for distances.)

The Gall-Peters projection is an equal-area projection that is like the Lambert equal-area cylindrical except that horizontal coordinates are multiplied by 0.5 so that shapes are good at $45°$ latitude rather than at the equator. This was discovered first by Gall and later independently by Peters. Africa is too thin and the polar areas are too fat. Peters famously promoted this as the fairest projection of the earth to third world countries because of its equal-area feature, but it was rightly criticized by cartographers because other equal-area projections with smaller isotropy errors such as the Hammer, Mollweide, Breisemeister, and Eckert IV were already known. The Gall-Peters projection is pretty good at distances ($\sigma = 0.390$) (better than Mercator), but it would be better still if the horizontal coordinates in the Lambert Cylindrical equal-area horizontal coordinates were multiplied by 0.4 instead of 0.5. This would give distance errors of $\sigma = 0.384$ but would have an even thinner Africa and would still not be particularly outstanding for distances. Cylindrical maps that minimize distance errors tend to be compressed in the horizontal direction (making Africa too thin) because they are trying to shrink the distances in the polar regions and also bring points on opposite sides of the date line closer together. Cylindrical maps therefore do not achieve as good distance errors as some other types.

Interestingly, and not surprisingly, the maps with the best distances turn out to all be nearly circular. The polyconic, despite its unusual shape does surprisingly well with $\sigma = 0.364$. This projection has constant distances on the prime meridian and along lines of latitude. Except for the pinching in at the poles it is an approximately circular map. An azimuthal projection centered at (0°,0°) would be even better. Azimuthal projections offer the possibility of being best overall. The Mollweide projection is an ellipse with a 2:1 axis ratio. The hemisphere centered on (0°,0°) is a circle while the other hemisphere is in two lobes to the left and right. Distances would obviously be improved if these two lobes were smeared out to make an azimuthal ring. The (azimuthal) stereographic projection (for the whole sphere) has distance errors of $\sigma = 0.714$ which is worse than the Mercator, but the equidistant azimuthal projection has distance errors of $\sigma = 0.356$, and the Lambert equal-area azimuthal projection has distance errors of only $\sigma = 0.343$ or slightly better than the Gott-Mugnolo equal-area elliptical.

## 4   Force Method for Improving Distances

We were interested to see if the Lambert equal-area azimuthal projection could be improved upon. Thus Gott and Mugnolo placed 3,000 particles down randomly on the sphere and mapped them onto the plane with the Lambert equal-area azimuthal projection. Then we established a radial force between each pair of particles on the map governed by a potential which is proportional to the square of the logarithmic distance error for that pair of particles on the map:

$$U = \tfrac{1}{2}\left[\ln\left(d_{\text{map}} / d_{\text{globe}}\right)\right]^2 \qquad (4)$$

Then the force between two of particles is $F = -dU/dr$ where the force is central and $r$ is the distance between the two particles on the map.

Each particle is then allowed to move under the vector sum of the forces from the other particles in the planar map. This is an $N$-body problem in the plane, just like a gravitational force except that it is not a $1/r^2$ force but a force governed by the potential above. Since energy is conserved, as the particles pick up kinetic energy the potential energy drops. After one time step, the particles' motion is stopped—resetting their kinetic energy to zero. The total potential energy of the system which is proportional to $\sigma^2$ for the map [the sum of the potential energies for each pair of random points] should be less than before. After each timestep, an overall scale factor is chosen to minimize $\sigma^2$. This is then repeated until the system settles into a relaxed distribution where the potential and therefore the rms logarithmic distance errors between pairs of points is in a local (and perhaps global) minimum. Timesteps are taken short enough to avoid overshooting of the equilibrium. The 3,000 particles give 4,498,500 pairs of distances. The radial distribution of points ($r$ as a function of latitude $\phi$) can then be plotted. This makes a tight scatter diagram which can be approximated by the simple analytic formula:

$$r = \sin\left[0.446\left(\frac{\pi}{2} - \phi\right)\right] \qquad (5)$$

(c.f. Mugnolo 2005 for details). The formula for the Lambert equal-area azimuthal is similar except "0.446" is replaced by "0.5". Thus our formulae for the ($x,y$) Cartesian coordinates on the Gott-Mugnolo azimuthal projection centered on the north pole are:

$$x = \cos \lambda \sin\left[ 0.446\left(\frac{\pi}{2} - \phi\right) \right]$$

$$y = \sin \lambda \sin\left[ 0.446\left(\frac{\pi}{2} - \phi\right) \right]$$

(6)

Checking with 30,000,000 random pairs, we find that this map projection has distance errors of only $\sigma = 0.341$ which is the lowest of all the map projections we have studied. The value of the optimal constant 0.446 was then checked to an accuracy of two significant figures by varying this constant and minimizing the errors. Given the symmetry of the problem, it is perhaps not surprising that the best projection of the sphere for distance errors gives a map that is circular in shape. The Gott-Mugnulo Azimuthal projection has smaller isotropy and skewness errors (Goldberg & Gott 2006) than the Lambert equal-area azimuthal and better area errors than the Equidistant Azimuthal, which gives only $1/4^{th}$ of the area to the central and best portrayed hemisphere.

A map of the Earth on the Gott-Mugnolo projection centered at $(0^o, 0^o)$ is shown in Figure 4. This might be appropriate for showing the history of colonization of our species, homo sapiens, over the globe. Starting in Africa 200,000 years ago, we went into Eurasia about 100,000 years ago, then into Australia 60,000 years ago, and then into North and South America and finally into Antarctica (the first people born in Antarctica had parents from Argentina). Here it is important to avoid boundary cuts because one wants to chart the movement across the Bering Straight land bridge from Asia into North America without interruption. Also, distances traveled are important, not just from Africa, but between continents. We want to show all continents whole and uninterrupted. So a map centered on Africa, minimizing distances, and without boundary cuts is ideal.

The Gott-Mugnolo projection is also useful for a map of the entire moon that is centered on the face we observe from earth. See Figure 5. This image has been composed from Clementine satellite images (NRL 2006). The projection portrays the central hemisphere well (the face of the moon we are used to). The north pole and the south pole appear as spots where there are significant shadows. This montage is produced using images that have the sun as nearly overhead locally as possible, but the north pole and south pole always see the sun at low angle, so shadows predominate. The circle centered on the center of the map that includes the north and south pole contains the face of the moon we see from earth. The map projection allows you to look at the face of the moon we usually see and yet peek around the edges and see the back side of the moon that we never see from earth. Most of the dark "seas" (old lava flows actually) on the moon are on the face toward the earth. On the left at about 8 o'clock, the leftmost small sea is Mare Imbrium, due to a very old large asteroid impact. This is on the left limb of the moon (we only see half of it very foreshortened from the earth). We can see it much better on this map. On the right, at about 4 o'clock, the small dark sea Tsiolkowski, named after the famous Russian space visionary, can be seen. This was discovered by the first Russian probe to photograph the back side of the moon.

In a separate paper (Gott, et al. 2006), we are showing the WMAP cosmic microwave background 3 year map of the sky in both the Gott equal area Elliptical and the Gott-Mugnolo Azimuthal projections.

As computers become faster, it will be possible to repeat this calculation with more than 3,000 points. This would allow a tighter scatter diagram and may make it possible to find a more complicated analytic fitting formula which would be more accurate, but we expect that our current analytic approximation should produce logarithmic rms distance errors that are good enough to three significant figures. And it is an advantage to have a simple mathematical formula for map making.

In the future, the method could be applied to make distance minimizing maps for continents and smaller regions. One could make a distance minimizing map of the United States for example by placing random points inside the US boundaries on the globe, and then moving these with the force law until equilibrium and the best map was achieved. One could start with a Lambert azimuthal equal-area projection centered on the geographic center of the US and then move the points on the map. Alaska and Hawaii would move into their optimal positions and achieve good shapes. Other points, like points on the coastline of Canada, would be massless points that exert no force, but would be attracted to the random points by the force law. If a point on the Canadian coastline is too close to points in the US it will be repelled, if it is too far away it will be attracted. It will eventually settle at a location that minimizes its rms logarithmic distance errors with respect to the random points in the US. The massless particles can be moved along with the massive particles as they seek equilibrium. Another alternative would be to replace the random points in the US with points on a fine mesh longitude and latitude grid such that their distribution is equal area [i.e., points ($\phi,\lambda$) where $n$ and $m$ are integers and $\sin\phi = -1 + 2(n/N)$, $\lambda = 2\pi(m/M)$ and $0 < n < N$, and $0 < m < M$, and $N$ and $M$ are large integers]. Then the points to be moved are on a grid within the borders of the US and cities and towns inside the US can be plotted by interpolation. Various schemes could be employed.

For world maps, one could have massive random points located only in land areas. This would produce a map with the minimal distance errors between points on land. If one wanted a map that would minimize the airline distance errors between the world's 50 largest cities, then one would have 50 massive points (the cities) that would move under the force law, until they reached equilibrium, and many massless points on continental coastlines that would be attracted or repelled by the force exerted by the massive points on them, and they would move until they reached equilibrium. We have tried some experiments in this regard that look promising (see Mugnolo 2005 for more details). In the limit where one had only 2 cities, this process would lead to the 2-pt equidistant projection where the distance between the 2 cities is exact and the distances of any other point on the globe to the two cities is also exact.

The force law technique operating on random points could also be used on irregular bodies (like asteroids) to produce the best possible distance maps. It could also be used to produce the best possible 2 or 3 dimensional maps of higher dimensional objects like polyhedroids, where the vertices would be the massive points. Or it could be used to plot 2D diagrams of points in multidimensional spaces where distances between key points are represented as well as possible.

## 5  Conclusions

We have developed a distance error criterion that allows us to compare different map projections. Since a distance error of a factor two too small between points is just as bad as an error of a factor of two too large, it is the rms logarithmic distance errors between random points on the globe, $\sigma$, that we will be trying to minimize. Most projections have better distance errors than the Mercator ($\sigma = 0.444$). We have also developed some new map projections that minimize distance errors. Map projections that do particularly well on distances are: the Breisemeister ($\sigma = 0.372$), Gott equal-area Elliptical ($\sigma = 0.365$), Polyconic ($\sigma = 0.364$), Azimuthal Equidistant ($\sigma = 0.356$), Gott-Mugnolo equal-area Elliptical ($\sigma = 0.348$), Lambert equal-area Azimuthal ($\sigma = 0.343$), Gott-Mugnolo Azimuthal ($\sigma = 0.341$). The Gott-Mugnolo Azimuthal which we introduce here is the best projection for distance errors. It was constructed using a new force method between pairs of points which can be used in principle to make distance error minimizing maps of smaller regions. One of the key features that maps try to produce is accurate distances between points. If one wanted to use a single measure to rate map projections then distance error would be a good candidate, since it reflects all kinds of local and large scale distortions in maps. However for general use we may wish to consider a number different measures and weight them appropriately. In the next paper (Goldberg & Gott 2006) we will consider isotropy, area, flexion (bending), skewness (lopsidedness), distance, and boundary cuts. Traditionally map projections have been judged and evaluated primarily on local isotropy and area errors as measured by Tissot (1881) ellipses. In the next paper we will add to these new measures for flexion, skewness and boundary cuts, as well as the distance criterion developed here to evaluate map projections. In that paper we will show various map projections with new indicatrices that illustrate the new measures for flexion and skewness as well as isotropy and area. Our force moving techniques may find future applications for maps of particular regions, as well as for the mapping of irregularly shaped asteroids. Also of interest, in a separate paper we have prepared a conformal map of the universe based on the logarithm map of the complex plane (Gott, et al. 2005).

Thanks to Mario Juric for preparing one of the world map projection diagrams for us, and to David Goldberg for calculating the distance errors for several additional map projections we added at the end of our study, and for many helpful conversations. JRG was supported by NSF Grant AST04-06713. WNC thanks the Center for Modeling, Simulation and Analysis for providing computational support for his work.

## 6 Figures and Tables

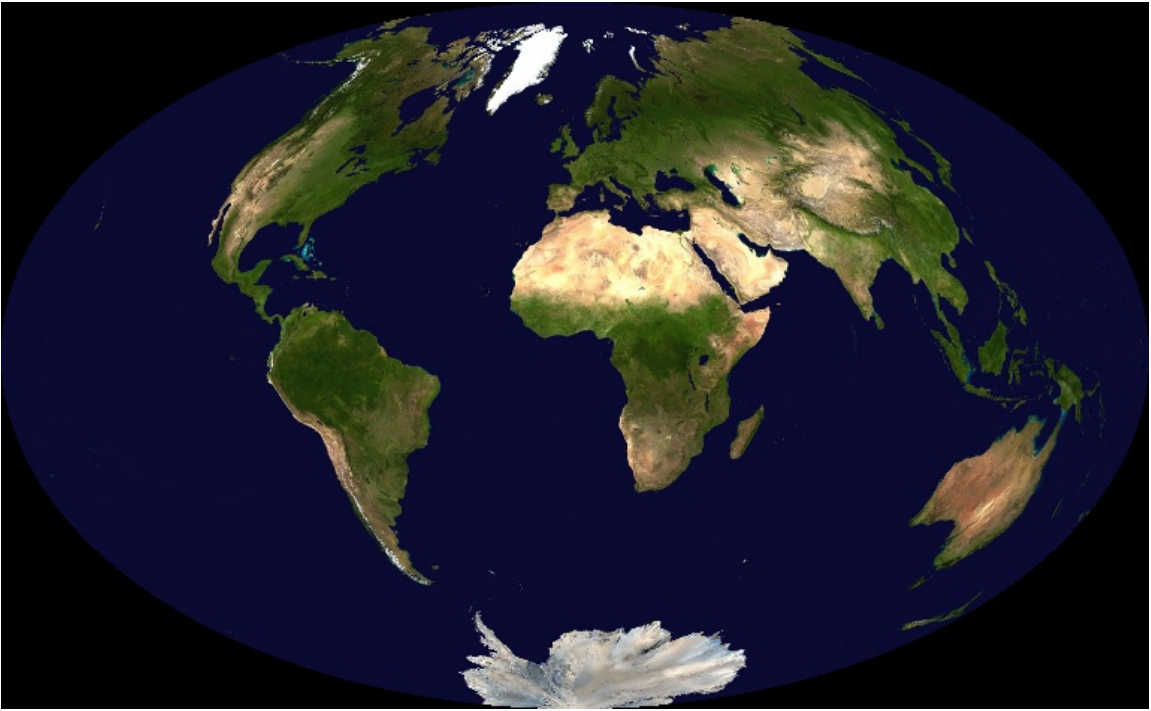

**Figure 1** *The Gott equal-area elliptical projection of the Earth. Shapes are perfect locally along the Greenwich Meridian. The distance scale is also linear along this meridian. This has good shapes for Europe, Africa, and Antarctica. The polar areas are better displayed than in the Mollweide projection. Since the map is more nearly circular than the Mollweide it makes smaller distance errors for points on opposite sides of the international date line in the Pacific. Also, the lengths of the different meridians are more nearly equal on the map. This has rms logarithmic distance errors of $\sigma = 0.365$. Source image: Visible Earth (2006).*

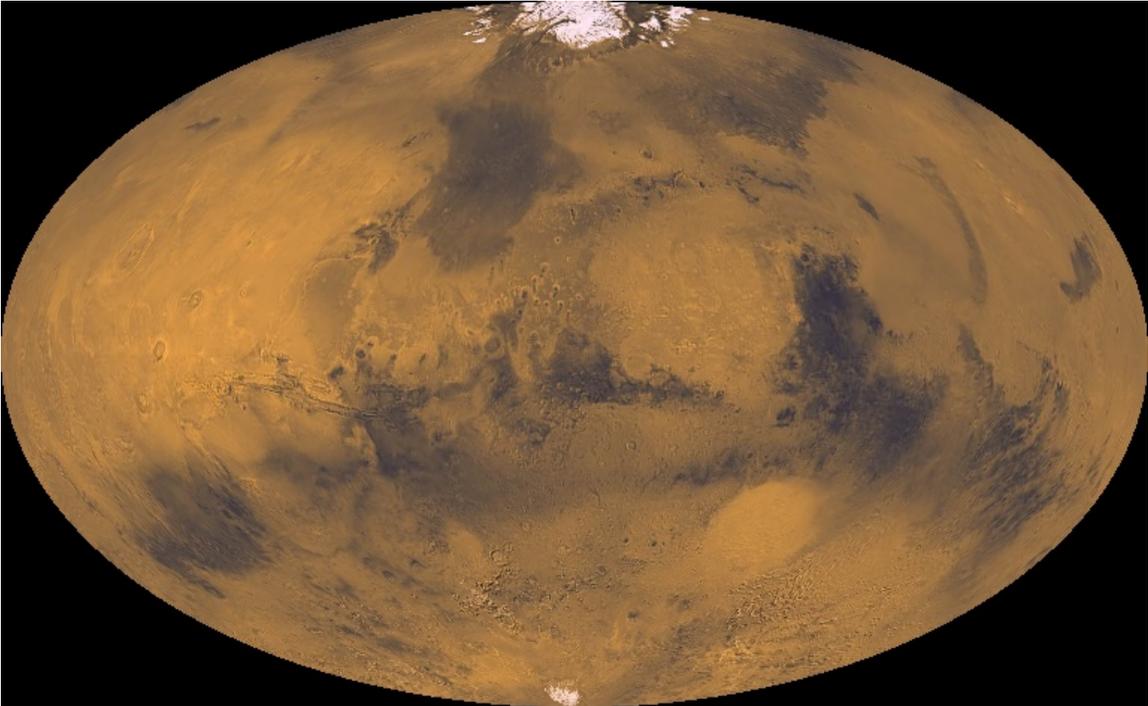

**Figure 2** *The Gott equal-area elliptical projection of Mars. The polar caps are portrayed well, as is Syrtis Major, the most prominent dark feature on the planet (dark region looking a little like Africa upside down), and Hellas, the bright red circular region below it and slightly to the left. Source image: JPL (2006).*

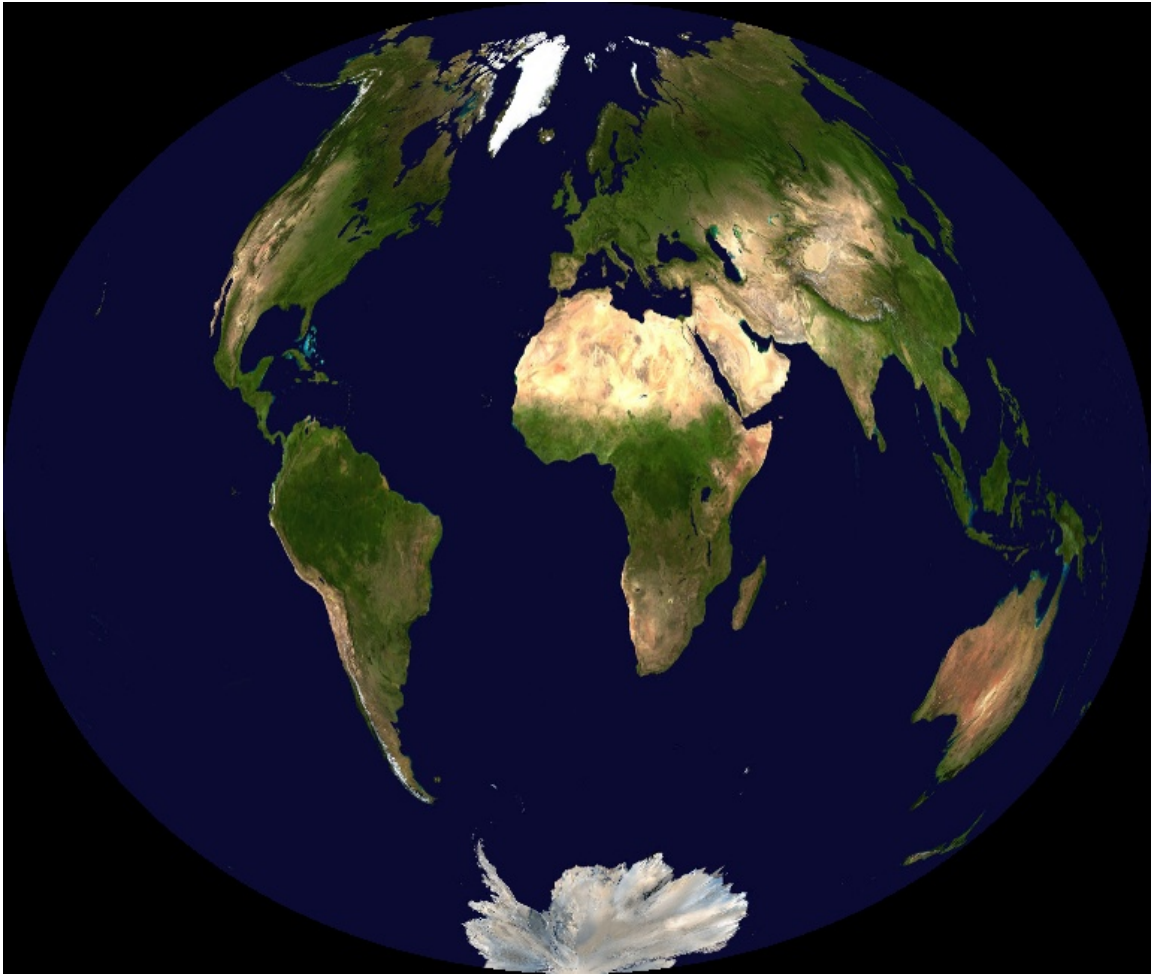

**Figure 3**  *The Gott-Mugnolo equal-area elliptical projection with $\sigma = 0.348$, the smallest distance errors of any map projection we have studied with the north pole at the top of the map and the south pole at the bottom. This is a compressed version of the Gott projection in figure 1.  Africa, and Europe are now have shapes that are too thin, but the map is more nearly circular, so the meridians are of more nearly equal length, and points on opposite sides of the date line in the Pacific are now closer together so their distance errors are smaller.  Source image: Visible Earth (2006).*

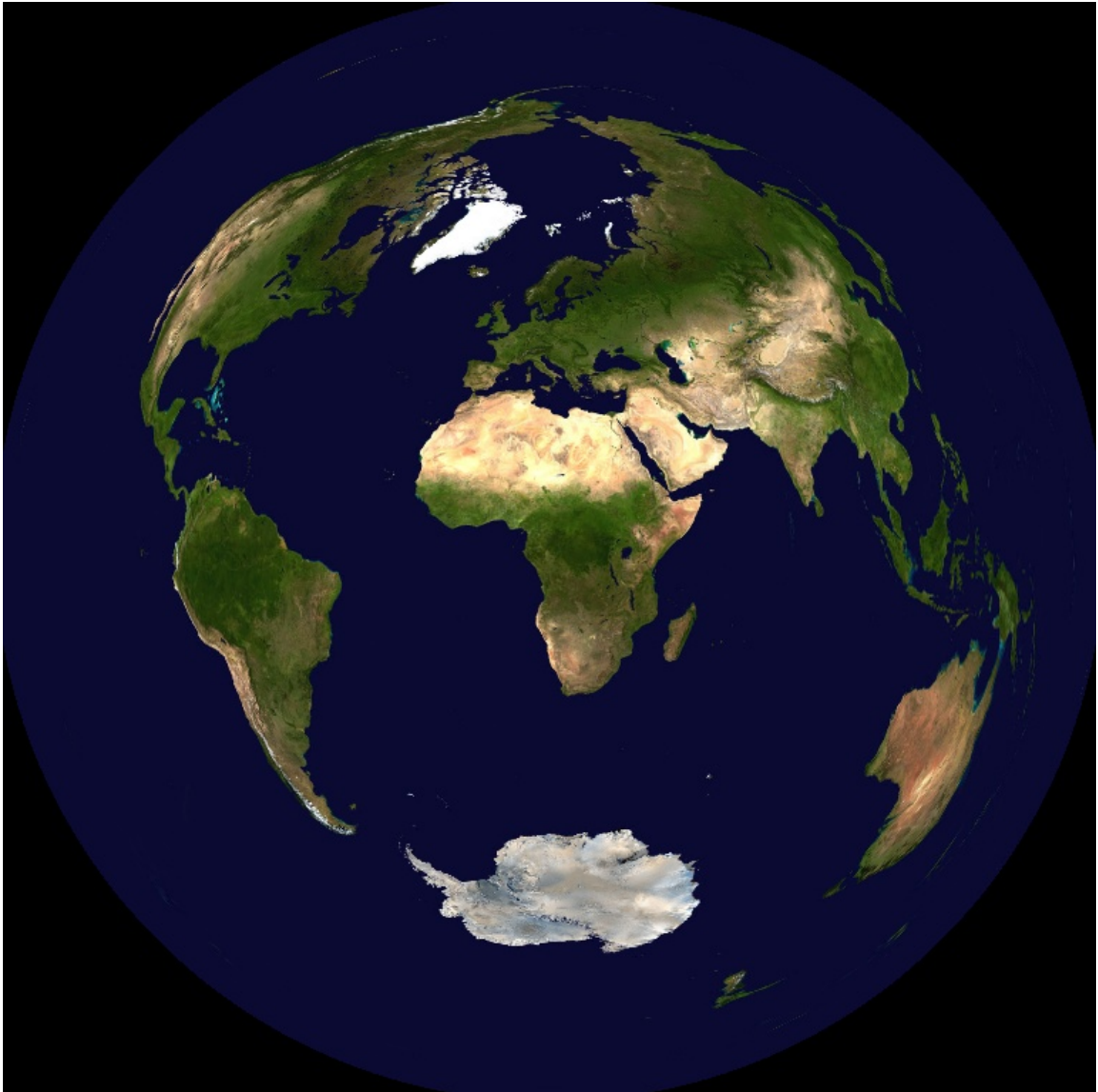

**Figure 4** *The Gott-Mugnolo azimuthal projection centered on 0° latitude, 15° east longitude. It has distance errors of only $\sigma = 0.341$ which is the best of any projection we have studied so far. Like the Lambert azimuthal, it gives an excellent portrayal of the central hemisphere whereas the other hemisphere has rather distorted shapes. But it has smaller local distortions in the opposite hemisphere than the Lambert equal-area azimuthal. Antarctica and Australia are less squashed in this projection than they would be in the Lambert equal-area azimuthal. The configuration of continents looks rather similar to that in figure 3, except that Antarctica is seen as approximately elliptical in shape with the south pole in the center, rather than at the bottom of the map. Africa is not stretched. Source image: Visible Earth (2006).*

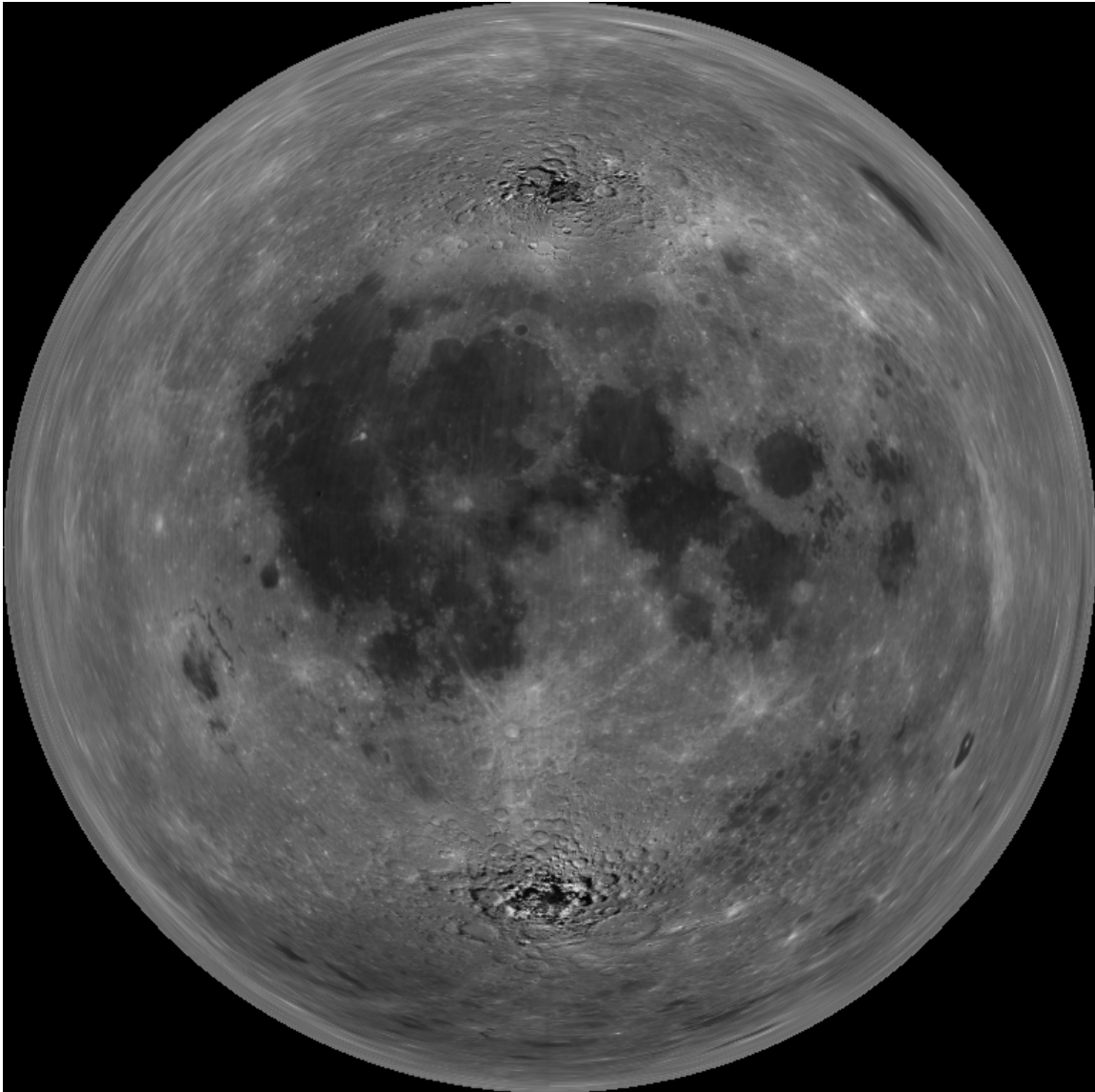

**Figure 5** *The Gott-Mugnolo azimuthal projection for the Moon centered on the face we see from the earth. The face we see from earth is circular, centered on the center of the map and with a radius 65.4% of the radius of the map. We can see the familiar seas visible from earth in the center of the map, while we are allowed to peek around the edge to see the far side as well. The north and south pole (at the edge of the visible face), which always see the sun at low illumination angle are the two points in the map that show strong shadowing of craters. Mare Imbrium can be seen on the left at about 8 o'clock, while Tsiolkowski can be seen on the right at about 4 o'clock. This map has been constructed from Clementine satellite photos. Source image: NRL (2006).*

**Table 1:** *World maps' rms logarithmic distance errors.*

| Projection | RMS Logarithmic Distance Error |
|---|---|
| Gott-Mugnolo Azimuthal | 0.341 |
| Lambert Azimuthal (EA) | 0.343 |
| Gott-Mugnolo Elliptical (EA) | 0.348 |
| Azimuthal Equidistant | 0.356 |
| Polyconic | 0.364 |
| Gott Elliptical (EA) | 0.365 |
| Breisemeister (EA) | 0.372 |
| Eckert VI (EA) | 0.385 |
| Hammer (EA) | 0.388 |
| Aitoff | 0.389 |
| Eckert IV (EA) | 0.390 |
| Mollweide (EA) | 0.390 |
| Gall Isographic | 0.390 |
| Gall-Peters (EA) | 0.390 |
| Kavrayskiy VII | 0.405 |
| Winkle-Tripel | 0.412 |
| Orthographic Lagrange | 0.414 |
| Gall-Stereographic | 0.420 |
| Bromley-Mollweide (EA) | 0.420 |
| Lambert Azimuthal (2-Hem) (EA) | 0.425 |
| Lagrange (C) | 0.432 |
| Azimuthal Equidistant (2-Hem) | 0.432 |
| Miller | 0.439 |
| Braun Stereographic | 0.441 |
| Mercator (C) | 0.444 |
| Equirectangular | 0.449 |
| Lambert Conic | 0.460 |
| Lambert Cylindrical (EA) | 0.473 |
| Mod. Braun Stereographic | 0.627 |
| Stereographic (2-Hem) (C) | 0.692 |
| Stereographic (C) | 0.714 |

**Table 2:** *Hemisphere Maps: rms logarithmic distance errors.*

| Projection | RMS Logarithmic Distance Error |
|---|---|
| Lambert Azimuthal (EA) | 0.0829 |
| Azimuthal Equidistant | 0.0844 |
| Stereographic (C) | 0.127 |
| Orthographic | 0.164 |
| Gnomonic | 1.090 |